\documentclass[showpacs,aps,prc,longbibliography,nofootinbib,showkeys,superscriptaddress,twocolumn,longbibliography]{revtex4-2}
\usepackage{graphicx}
\usepackage{bm}
\usepackage{multirow}
\usepackage{natbib}
\usepackage{array}
\usepackage{amsmath}
\usepackage{mathrsfs}
\usepackage{amssymb}
\usepackage{float}
\usepackage{xcolor}
\usepackage{enumerate}
\usepackage{color}
\usepackage{slashed}
\usepackage[colorlinks]{hyperref}
\usepackage{url}
\usepackage{txfonts}
\usepackage{booktabs}
\usepackage{BOONDOX-cal}
\usepackage{orcidlink}
\usepackage{longtable}
\usepackage{makecell}
\usepackage{cancel}
\usepackage{cleveref}

\usepackage{tikz}
\usetikzlibrary{arrows.meta}
\usetikzlibrary{decorations.pathmorphing, decorations.pathreplacing, calc}
\usetikzlibrary{fadings}
\usetikzlibrary{decorations.markings}

\allowdisplaybreaks[4]
\hypersetup{hypertex=true,colorlinks=true,linkcolor=blue,anchorcolor=red,citecolor=blue}

\newcommand{\feq}{f_\mathrm{eq}}
\newcommand{\dif}{\mathrm{d}}

\begin{document}
\title{Fermi-liquid view of viscosity in cold and dense nucleon matter}
\author{Jianing Li\orcidlink{0000-0001-7193-7237}}\email{ljianing@physi.uni-heidelberg.de}
\affiliation{Physikalisches Institut, Universit\"at Heidelberg, 69120 Heidelberg, Germany}
\affiliation{GSI Helmholtzzentrum f\"ur Schwerionenforschung, 64291 Darmstadt, Germany}
\affiliation{Key Laboratory of Quark and Lepton Physics (MOE) and Institute of Particle Physics, Central China Normal University, Wuhan 430079, China}
\author{Weiyao Ke\orcidlink{0000-0002-5630-2388}}\email{weiyaoke@ccnu.edu.cn}
\affiliation{Key Laboratory of Quark and Lepton Physics (MOE) and Institute of Particle Physics, Central China Normal University, Wuhan 430079, China}
\author{Jin Hu\orcidlink{0000-0003-1179-4603}}\email{hu-j23@fzu.edu.cn}
\affiliation{Department of Physics, Fuzhou University, Fuzhou 350116, China}
\date{\today}
\begin{abstract} 
We develop a framework to calculate transport properties in cold, dense relativistic quasiparticle system within the Fermi-liquid theory at the mean-field level. 
Building on our previous study J. Li \emph{et al.} [\href{https://doi.org/10.1103/PhysRevC.111.044904}{Phys.\ Rev.\ C \textbf{111}, 044904 (2025)}], we start from the linearized relativistic Boltzmann equation tailored to quasiparticles with medium-dependent dispersion relation and implement Landau matching conditions, proving that the bulk viscosity is manifestly nonnegative. A low-temperature expansion then yields leading-order ($T/\mu^*$) expressions for the shear ($\eta$) and bulk ($\zeta$) viscosities, where the behavior $\zeta/\eta \propto (T/\mu^*)^4$ in the degenerate regime is found to be robust against quasiparticle mass correction. 
We couple the kinetic framework to a Walecka-type mean-field equation of state and compute $\eta$ and $\zeta$ for cold, dense nucleon matter.
The transport properties of nucleonic matter in the degenerate regime can be relevant for intermediate beam-energy nuclear experiments.
\end{abstract}
\maketitle
\section{Introduction}
The transport phenomena is of vital importance in the study of physical systems out of equilibrium, and often show universal behaviors across orders of magnitude --- from the microscopic relaxation in nuclear collisions to the dissipation processes in macroscopic astrophysical systems~\cite{de2013non,Botermans:1990qi,Shapiro:1983du}. Since the discovery of the color-deconfined matter of quantum chromodynamics (QCD) in ultrarelativistic heavy-ion collisions~\cite{PHENIX:2004vcz,ALICE:2010suc}, the precise characterization of its transport properties has always been a central focus~\cite{Heinz:2013th}. 
The understanding of the non-equilibrium process plays indispensable roles in identifying the partonic degrees of freedom, extracting shear and bulk viscosity, and constraining the QCD equation of state of the hot QCD matter from data~\cite{MUSES:2023hyz}. Many developments of the non-equilibrium theory to study the hot QCD rely on the predominant temperature as the solely relevant energy scale that determines various microscopic time scales~\cite{Baier:2000sb,Arnold:2002zm,Romatschke:2017ejr,Berges:2020fwq}.
In contrast, transport inside cold and dense nucleonic matter is governed by quasiparticle excitations near a sharp Fermi surface, set by the Fermi energy (or Fermi momentum), and dissipation involving hadronic interactions among the these particles~\cite{ring2004nuclear}. 
A controlled description of nucleonic transport is now especially timely. 

The compression and excitation of nuclear matter at low temperature relative to the baryon chemical potential $T\ll \mu_B$ can be achieved in intermediate-energy heavy-ion facilities~\cite{Yang:2013yeb,Senger:2016wfb,Bzdak:2019pkr,Agarwal:2023otg}, creating a unique environment to study its transport properties and to pursue hydrodynamic and other collective behavior in this regime. Furthermore, the search for collective effects has expanded to include ultra-peripheral collisions, which also require a precise understanding of transport at high baryon density~\cite{Zhao:2022ayk, Shen:2022oyg, Schenke:2024cnc}. Upcoming facilities like the Electron-Ion Collider (EIC) in the United States~\cite{Accardi:2012qut} and the proposed Electron-Ion Collider in China (EicC)~\cite{Anderle:2021wcy} promise to probe this cold, dense matter in a new way, potentially allowing us to observe the propagation of shocks through it. However, interpreting the resulting signals, which are sensitive to density gradients, compressional flows, and entropy production, presents a significant challenge. Similar mechanisms are probed in other low-energy nuclear experiments where the system remains far from deconfinement yet exhibits nonequilibrium dynamics on hadronic scales~\cite{Aichelin:1991xy,harakeh2001giant,Galatyuk:2014vha}. These considerations underscore the need for a dedicated framework for nucleonic transport at low temperature. It requires transport coefficients, specifically the shear and bulk viscosity ($\eta$ and $\zeta$), that are not merely phenomenological parameters but are fundamentally consistent with the symmetries, thermodynamics, and microscopic interactions of nuclear matter.

Such cold/warm and dense nuclear matter is also realized in neutron stars and neutron star merger events, where its transport properties play an important role in stellar cooling~\cite{Douchin:2001sv,Schmitt:2017efp,Shternin:2020igy,Tsang:2023vhh}. In that context, however, dissipation is dominated by slow electroweak processes~\cite{Hoyos:2020hmq,CruzRojas:2024etx}. This work aims to establish the essential link between transport related to nuclear experimental observables and the intrinsic properties of cold, dense nuclear matter. In particular, we focus on strong-interaction processes acting on short time scales~\cite{sakurai1960theory}, where many-body scattering near the Fermi surface governs the attenuation of collective modes~\cite{Delacretaz:2022ocm,Li:2024iqt}.

In our previous work~\cite{Li:2024iqt}, we analyzed cold, dense nucleon matter within a mean-field Fermi-liquid framework and computed transport coefficients in the Walecka model~\cite{Walecka:1974qa}. That study revealed a gap in the understanding: In a generic quasiparticle picture, the sign of $\zeta$ is not automatically guaranteed positive without further constraints. This is because, in our previous analysis, an appropriate matching condition for such a quasiparticle fermionic system was not yet fully established. In the present work, we remedy this limitation by establishing consistency between transport theory and hydrodynamics. Specifically, we develop a general, self-consistent scheme for dense and cold quasiparticle systems in which $\zeta$ is rigorously non-negative under transparent assumptions tied to thermodynamic consistency and microscopic reversibility. Within this framework, the symmetry constraints of the linearized collision kernel, and their restoration within the relaxation-time approximation (RTA) treatment, are encoded in a set of Landau matching conditions. A direct implication is that in cold Fermi systems, $\zeta$ is strongly suppressed at low temperature, rendering bulk dissipation parametrically smaller than shear dissipation. We can conclude that the effect of mean-field dressed mass will not alter the general behavior of Fermi-liquid that $\zeta/\eta \propto (T/\mu^*)^4$ at low temperature.

This paper is organized as follows. In Sec.~\ref{sec1.1} we formulate the linearized relativistic Boltzmann equation for a Fermi liquid of quasiparticles and derive the expressions for $\eta$ and $\zeta$ within the RTA. In Sec.~\ref{sec1.2} we implement Landau matching conditions adapted to quasiparticles and prove a general positivity result for $\zeta$, while Sec.~\ref{sec1.3} develops the low-temperature expansion scheme. In Sec.~\ref{sec2} we apply this framework to a Walecka-type mean-field equation of state and present numerical results for cold, dense nucleon matter. Section~\ref{sec3} summarizes the main conclusions and discusses implications for phenomenology and future directions.
\section{Transport in Fermi-liquid}\label{sec1}
\subsection{Linearized relativistic Boltzmann equation framework}\label{sec1.1}
To investigate the transport properties of a quasifermion system, we begin with the linearized relativistic Boltzmann equation,  
\begin{align}
	\label{eq:Boltzmann_equation}
	\left(\frac{\partial}{\partial{t}}+\frac{\boldsymbol{p}}{E_{\boldsymbol{p}}^*}\cdot\nabla_{\boldsymbol{x}}-\nabla_{\boldsymbol{x}}E_{\boldsymbol{p}}^*\cdot\nabla_{\boldsymbol{p}}\right)f_{\rm eq}\left(x,p^*\right)=\mathcal{C}\left[\delta{f}\right]\,.
\end{align}
Here $f_{\rm eq}(x,p^*)$ denotes the local equilibrium distribution function, and
$\mathcal{C}[\delta f]$ is the corresponding linearized collision kernel, obtained by expanding the full
nonlinear collision kernel $\mathcal{C}_{\rm full}[f]$ to first order in the deviation $\delta f$ around
$f_{\rm eq}$. The nonequilibrium correction $\delta f(x,p^*)$ is defined by
\begin{align}\label{eq:definition_of_f}
    f(x,p^*) = f_{\rm eq}(x,p^*) + \delta f(x,p^*) \, .
\end{align}

The equilibrium distribution takes the Fermi-Dirac form, 
\footnote{Throughout this work, we consider a dense fermionic system in which particle contributions strongly dominate over antiparticles, antiparticle effects are therefore neglected.}
\begin{align}
	\feq\left(x,p^*\right)=\frac{1}{e^{(u\cdot p^*-\mu^*)/T}+1}\,,
\end{align}
where $T$ denotes the temperature, $u^\mu$ the fluid four-velocity, $\mu^*$ the effective chemical potential, i.e., the energy required to add one more particle to the local system. Because of repulsive part of the mean-field interactions, $\mu^*$ is different from $\mu$, with the latter being the thermodynamic conjugate to the particle number density.
The mean field interaction also modifies the quasiparticle dispersion relation to be $E_{\boldsymbol{p}}^*=\sqrt{{m^*}^2+\boldsymbol{p}^2}$ with effective mass $m^*$ and momentum $\boldsymbol{p}$, and the corresponding four-momentum is ${p^*}^\mu=\left(E_{\boldsymbol{p}}^*,\boldsymbol{p}\right)$.
Both $m^*$ and $\mu^*$ are functions of $\mu$.
 
The non-equilibrium correction associated with a deviation of $\mu$ can be assumed as
\begin{align}
\delta f = \tilde{f}_{\mathrm{eq}}\phi\,,
\end{align}
with
\begin{align}
	\tilde{f}_\mathrm{eq}\left(x,p^*\right)\equiv\frac{\partial\feq\left(x,p^*\right)}{\partial\mu^*}\,.
\end{align}
The function $\phi$ can be decomposed to pieces containing different tensor structure of the flow grdients
\begin{align}
	\label{eq:phi}
	\phi=\phi_1{p^*}^\mu {p^*}^\nu\sigma_{\mu\nu}-\phi_2\vartheta\,,
\end{align}
with
\begin{align}
    \begin{split}
    \sigma_{\mu\nu}&=D_\mu u_\nu+D_\nu u_\mu-\frac{2}{3}\Delta_{\mu\nu}\vartheta\,,\\
    \Delta^{\mu\nu}&=g^{\mu\nu}-u^\mu u^\nu\,,\\
    D_\mu&=\Delta_{\mu\nu}\partial^\nu\,,\\
    \vartheta&=\partial\cdot{u}\,,
    \end{split}
\end{align}
where we have used the mostly-plus metric $g^{\mu\nu}=\mathrm{diag}(1,-1,-1,-1)$. 

The energy-momentum tensor of the system can be expanded near thermal equilibrium
\begin{align}
	\label{eq:Tmunu0}
	T^{\mu\nu}=-pg^{\mu\nu}+\left(p+\varepsilon\right)u^\mu u^\nu+\delta{T}^{\mu\nu}_\mathrm{diss}\,,
\end{align}
where $p$ and $\varepsilon$ denote pressure and energy density, respectively. To the first order in gradients, the dissipative correction is
\begin{align}
	\label{eq:disspative part in Tmunu1}
	\delta{T}^{\mu\nu}_\mathrm{diss}=\eta\sigma^{\mu\nu}-\zeta\Delta^{\mu\nu}\vartheta\,,
\end{align}
with $\eta$ and $\zeta$ representing shear and bulk viscosities. In kinetic theory the same tensor can be expressed with the distribution function of the quasiparticles
\begin{align}
	\label{eq:Tmunu2}
	T^{\mu\nu}&=\int_{\boldsymbol{p}}\frac{{p^*}^\mu {p^*}^\nu}{E^*_{\boldsymbol{p}}}f\left(x,p^*\right) \nonumber\\
    &=\int_{\boldsymbol{p}}\frac{{p^*}^\mu {p^*}^\nu}{E^*_{\boldsymbol{p}}}f_{\rm eq}\left(x,p^*\right) + \int_{\boldsymbol{p}}\frac{{p^*}^\mu {p^*}^\nu}{E^*_{\boldsymbol{p}}}\delta f\left(x,p^*\right)\,,
\end{align}
with $\int_{\boldsymbol{p}}=\int\dif^3\boldsymbol{p}/\left(2\pi\right)^3$. The first term gives the equilibrium part of the energy-momentum tensor, then the dissipative part follows:
\begin{align}
	\label{eq:disspative part in Tmunu2}
	\delta T^{\mu\nu}_\mathrm{diss}=\int_{\boldsymbol{p}}\frac{{p^*}^\mu {p^*}^\nu}{E^*_{\boldsymbol{p}}}\delta f\left(x,p^*\right)\,.
\end{align}
Combining Eq.~\eqref{eq:disspative part in Tmunu2} with Eq.~\eqref{eq:phi} in local rest frame, one obtains the expression for bulk viscosity in kinetic theory
\begin{align}
	\label{eq:viscosity1}
	\eta=\frac{2}{15}\int_{\boldsymbol{p}}\frac{\boldsymbol{p}^4}{E_{\boldsymbol{p}}^*}\phi_1\tilde{f}_\mathrm{eq}\,,\quad
	\zeta=\frac{1}{3}\int_{\boldsymbol{p}}\frac{\boldsymbol{p}^2}{E_{\boldsymbol{p}}^*}\phi_2\tilde{f}_\mathrm{eq}\,.
\end{align}

\subsection{Quasiparticle matching condition for the relaxation-time approximation}\label{sec1.2}
In principle, one can obtain the viscosities $\eta$ and $\zeta$ by inverting Eq.~(\ref{eq:Boltzmann_equation}) to solve for $\delta f$, thus determining the perturbations $\phi_1$ and $\phi_2$. However, the linearized version of any valid collision operator should possesses the following properties
\begin{align}\label{eq:zero_mode}
\mathcal{C}\left[a + b^\mu p^*_\mu\right] = 0\,,
\end{align}
for arbitrary constant number $a$ and vector $b^\mu$.
They are consequences of particle-number and energy-momentum conservation of the microscopic two-body collision processes~\cite{groot1980relativistic}. Consequently, $1$ and $p^{*\mu}$ constitute zero modes of the collision operator. Therefore, we can only invert Eq.~(\ref{eq:Boltzmann_equation}) up to an arbitrary linear combination of the zero modes.
This means that the same out-of-equilibrium $T^{\mu\nu}$ can be decomposed in multiple ways into equilibrium and dissipative contributions, i.e., solutions of $\{\phi_1,\phi_2\}$ to Eq. (\ref{eq:Boltzmann_equation}) are not necessarily unique unless a convention is chosen to fix the state. 
The Landau matching conditions provide precisely such a prescription. They require that deviations of the local thermodynamic quantities do not alter the local energy-momentum density or conserved charge density $n^\mu$, i.e., 
\begin{align}\label{eq:LL_conditions}
    u_\nu u_\mu\delta T^{\mu\nu}=0\,,\quad
    u_\mu\delta n^\mu=0\,.
\end{align}
Correspondingly, they also define the flow velocity $u^\mu$ via the energy current $u_\nu T^{\mu\nu}=\varepsilon\,u^\mu$~\cite{landau1987fluid}.

For cold and dense fermionic matter we assume that $\delta f$ arises solely from spacetime variations of $\mu$, while fluctuations of $T$ are neglected. Within the RTA, the collision term is simplified as $\mathcal{C}\!\left[\delta f\right]=-\delta f/\tau_\mathrm{rel}$ with relaxation time $\tau_\mathrm{rel}$. \footnote{The RTA does not by itself preserve the zero modes of the linearized collision operator,  i.e., breaking Eq. \eqref{eq:zero_mode}. In practice, this violation is removed by imposing the matching conditions. Equivalently, one may view this as adding counterterms that ensure the collision operator annihilates the zero modes. The final results are therefore fully consistent with the symmetries of the microscopic theory. The implementation will be shown later.} The corresponding particular solutions for the coefficients are~\cite{Li:2024iqt}  
\footnote{At low temperature, the entropy contribution is negligible compared with the density effect, and is omitted during the formulation. Further details can be found in Ref. \cite{Li:2024iqt}.}
\begin{align}
	\begin{split}
		\phi_1\left(p^*\right)=\frac{\tau_\mathrm{rel}}{2E_{\boldsymbol{p}}^*}\,,\quad
		\phi_2^\mathrm{par}\left(p^*\right)=\frac{\alpha}{E_{\boldsymbol{p}}^*}\left[\left(u\cdot{p^*}\right)^2\gamma-{m^*}^2\right]\,.
	\end{split}
\end{align}
Here $v_s$ denotes the adiabatic speed of sound, and for later use we define
\begin{align}
	\alpha=\frac{\tau_\mathrm{rel}}{3}\,,\quad
	\gamma=1-3v_s^2\left(1-\frac{m^*}{\mu^*}\kappa^*\right)\,,\quad
	\kappa^*=\frac{\dif m^*}{\dif\mu^*}\,.
\end{align}
Note that $\phi_1$ is the unique solution, whereas $\phi_2^{\mathrm{par}}$ denotes a particular solution of $\phi_2$. In other words, the non-uniqueness induced by the zero modes is associated solely with $\zeta$. This is because $\sigma_{\mu\nu}$ is traceless and thus orthogonal to the conserved scalar quantities. In the local rest frame, owing to $\sigma_{0\nu}=0$ and $\delta^{ij}\sigma_{ij}=0$, the contribution proportional to $p^{i}p^{j}\sigma_{ij}$ in the isotropic system vanishes in the LL conditions. Thus, the shear term associated with $\phi_{1}$ always gives zero in the LL conditions, and the arbitrariness arises only in bulk one.

To obtain the full solution, we supplement the particular solution by deducting the zero-mode components $\left\{1,p^{*\mu}\right\}$ in the local rest frame \cite{groot1980relativistic},
\begin{align}
    \phi_2=\phi_2^{\mathrm{par}}-a-b E_{\boldsymbol{p}}^*\,.
\end{align}
The coefficients $a$ and $b$ are determined uniquely by the Landau matching conditions in Eq. \eqref{eq:LL_conditions}. The deviation of the total energy density becomes
\begin{align}
\delta{T^{00}}=\int_{\boldsymbol{p}}E_{\boldsymbol{p}}^*\delta{f}\,.
\end{align}
Similarly, the deviation of the total single-component conserved charge number density in the local rest frame is
\begin{align}
  \delta n^{0}
   = \int_{\boldsymbol{p}}\delta f\,.
\end{align}
For convenience, we define a inner product weighted by $\tilde f_\mathrm{eq}$ as
\begin{align}
	\langle X,Y\rangle\equiv\int_{\boldsymbol{p}}X\left(\boldsymbol{p}\right)Y\left(\boldsymbol{p}\right)\tilde{f}_\mathrm{eq}\,.
\end{align}
Then, the first Landau matching condition, enforcing energy density conservation in the local rest frame, reads
\begin{align}\label{eq:LL1}
	\langle E_{\boldsymbol{p}}^*,\phi_2\rangle=0\,,
\end{align}
The second matching condition, enforcing particle number conservation, gives
\begin{align}\label{eq:LL2}
	\langle 1,\phi_2\rangle=0\,,
\end{align}
Using Eq.~\eqref{eq:viscosity1}, the bulk viscosity can be expressed as
\begin{align}
	\zeta=\frac{1}{3}\langle \mathcal{R}_{\boldsymbol{p}},\phi_2\rangle\,,\qquad
    \mathcal{R}_{\boldsymbol{p}}\equiv\frac{\boldsymbol{p}^2}{E_{\boldsymbol{p}}^*}
    =\frac{{E_{\boldsymbol{p}}^*}^2-{m^*}^2}{E_{\boldsymbol{p}}^*}\,.
\end{align}
In general, $\phi_2$ admits the expansion $\phi_2 = \alpha \mathcal{R}_{\boldsymbol{p}}+\lambda_0 + \lambda_1 E_{\boldsymbol{p}}^*$. The Landau mathcing conditions imply that the coefficient vector $\boldsymbol{\lambda} = (\lambda_0, \lambda_1)^\mathrm{T}$ must satisfy
\begin{align}
	\label{eq:matrix_equation}
	\boldsymbol{H}\boldsymbol{\lambda} = -\alpha \boldsymbol{\mathcal{R}} ,
\end{align} 
with $H_{ij} \equiv \langle \psi_i, \psi_j \rangle$ and $\mathcal{R}_i = \langle \psi_i, \mathcal{R}_{\boldsymbol{p}} \rangle$ for $\psi_0=1$ and $\psi_1=E_{\boldsymbol{p}}^*$.
Since $\boldsymbol{H}$ is a Gram matrix of linearly independent variables, it is invertible~\cite{Horn_Johnson_2012}, and then Eq.~\eqref{eq:matrix_equation} always admits a solution for $\boldsymbol{\lambda}$. Combining this with the Landau matching conditions in Eqs. \eqref{eq:LL1} and \eqref{eq:LL2}, the bulk viscosity becomes
\begin{align}\label{eq:viscosity2}
    \zeta=\frac{1}{3\alpha}\Big\langle\left(\phi_2-\lambda_0-\lambda_1E_{\boldsymbol{p}}^*\right),\phi_2\Big\rangle
	=\frac{1}{3\alpha}\int_{\boldsymbol{p}}\tilde{f}_\mathrm{eq}\phi_2^2\geq0\,,
\end{align}
which is manifestly non-negative.

\begin{figure}[!htpb]
\centering
\resizebox{0.98\linewidth}{!}{
\begin{tikzpicture}[
  >=Stealth,
  font=\Large,
  every node/.style={align=center},
  state/.style={rounded corners=2pt, inner sep=6pt, outer sep=3pt, draw=black!20},
  lab/.style={inner sep=2pt, outer sep=2pt, text=black!85},
  note/.style={text=black!70},
  relabel/.style={-Stealth, ultra thick, draw=blue!70!black},
  noneq/.style={-Stealth, ultra thick, draw=orange!80!black},
  pseudo/.style={-Stealth, ultra thick, draw=black!50, dashed},
  muwave/.style={decorate, decoration={snake, amplitude=0.8mm, segment length=5mm}, line width=1.2pt}
]

\node[state] (f0)     at (0,0)      {$\feq\!\left(E_{\boldsymbol{p}}^*,\mu^*\right)$};
\node[state] (feqEeq) at (7,0)      {$\feq\!\left(E_{\boldsymbol{p}}^{\prime *},\mu^{\prime *}\right)$};
\node[state] (feqE)   at (14,0)     {$f\!\left(E_{\boldsymbol{p}}^{\prime *},\mu^{\prime *}\right)$};

\node[lab] (mu)  at ($(f0.north)+(0,1.5)$) {$\mu$};
\node[lab] (mup) at ($(feqE.north)+(0,1.5)$) {$\mu'=\mu+\delta\mu$};

\draw[muwave,->] (mu) -- (mup) node[midway, above=3pt, note] {$\delta\mu$};

\draw[relabel] (f0.east) -- (feqEeq.west)
  node[midway, above=3pt, note]{};

\draw[noneq] (feqEeq.east) -- (feqE.west)
  node[midway, above=3pt, note]{};

\draw[pseudo]
  ($(f0.east)+(0,-0.5)$) to[bend right=12]
  node[midway, below=3pt, note] {$\delta f$}
  ($(feqE.west)+(0,-0.5)$);

\end{tikzpicture}%
}
\caption{Separation of equilibrium relabeling induced by $\delta\mu$ from the nonequilibrium correction $\delta f$ obtained from the kinetic equation, where $E_{\boldsymbol{p}}^{\prime *}=\sqrt{m^*(\mu+\delta\mu)+\boldsymbol{p}^2}$ and $\mu^{\prime *}=\mu^*(\mu+\delta\mu)$.}
\label{fig:sketch}
\end{figure}
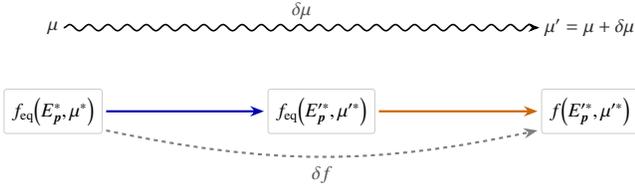

Finally, we note that the quasiparticle picture differs subtly from the particle picture. Since both $m^*$ and $\mu^*$ depend on $\mu$, an intermediate equilibrium state is established, as illustrated in Fig.~\ref{fig:sketch}. To maintain consistency with hydrodynamics, the Landau matching conditions still retain the same form as in the particle picture~\cite{Dusling:2013sea}. Some related discussions can also be found in Refs.~\cite{Chakraborty:2010fr,Albright:2015fpa}.
\subsection{Low temperature approximation}\label{sec1.3}
For later convenience, we define
\begin{align}
	\begin{split}
    \mathcal{S}_{-1}&\equiv \left\langle 1,\,1/E^{*}\right\rangle,\quad
		\mathcal{S}_0 \equiv \left\langle 1,\,1\right\rangle, \\
		\mathcal{S}_1&\equiv \langle 1\,, E^*\rangle,\quad
		\mathcal{S}_2 \equiv \langle E^*,\,E^*\rangle.
	\end{split}
\end{align}
The Landau matching conditions then yield
\begin{align}
	\begin{split}
		a &= \frac{\alpha {m^*}^2}{\Delta}\!\left(\mathcal{S}_2\mathcal{S}_{-1}-\mathcal{S}_1\mathcal{S}_0\right), \\
		b &= \alpha\gamma + \frac{\alpha {m^*}^2}{\Delta}\!\left(\mathcal{S}_0^2-\mathcal{S}_1\mathcal{S}_{-1}\right)\,,
	\end{split}
\end{align}
with $\Delta=\mathcal{S}_1^2-\mathcal{S}_2\mathcal{S}_0$.

We apply the Sommerfeld expansion to the equilibrium distribution in the local rest frame at low temperature \cite{Sommerfeld1928},  
\footnote{In the Sommerfeld expansion the lower integration limit is extended from $-\mu^*/T$ to $-\infty$, so the result is asymptotic rather than exact.}
%
\begin{align}
	\lim_{T\rightarrow0^+}\feq
	\approx \Theta\!\left(\mu^*-E^*\right)
	- \sum_{i=1}^{\infty}\mathcal{z}_i \frac{\dif^{2i-1}}{\dif {E^*}^{2i-1}} \delta\!\left(E^*-\mu^*\right),
\end{align}
with prefactor $\mathcal{z}_i = 2\left(1-2^{1-2i}\right) T^{2i}\zeta_{2i}$. Here, $\zeta_x \equiv \zeta(x)$ is the Riemann zeta function, $\Theta(x)$ the Heaviside function, and $\delta(x)$ the Dirac delta. Differentiating gives
\begin{align}
	\label{eq:tildefeq}
	\lim_{T\rightarrow0^+}\tilde{f}_\mathrm{eq}
	\approx \delta\!\left(\mu^*-E^*\right)
	+ \sum_{i=1}^{\infty}\mathcal{z}_i \frac{\dif^{2i}}{\dif {E^*}^{2i}} \delta\!\left(E^*-\mu^*\right),
\end{align}
in agreement with the canonical symmetry of the transport kernel in the Boltzmann equation at leading order (LO)~\cite{Delacretaz:2022ocm}. For LO, the viscosities are
\footnote{In the Sommerfeld expansion, because of the $\mu^*/T$ dependence in both the numerators and $\Delta$ through $a$ and $b$, the truncation order in Eq.~\eqref{eq:tildefeq} does not directly match the order of the viscosity results. Higher terms must be retained.}
%
\begin{align}\label{eq:LO_viscosity}
	\begin{split}
		\eta_\mathrm{LO} &= \frac{p_\mathrm{F}^{*\,5}}{30 \pi^2 \mu^*}\,\tau_\mathrm{rel}, \\
		\zeta_\mathrm{LO} &= \frac{8 \pi^2 m^{*\,4} p_\mathrm{F}^* T^4}{405\, \mu^{*\,5}}\,\tau_\mathrm{rel},
	\end{split}
\end{align}
so
\begin{align}\label{zeta_over_eta1}
	\frac{\zeta_\mathrm{LO}}{\eta_\mathrm{LO}}
	\propto \left(\frac{T}{\mu^*}\right)^4.
\end{align}
Thus, apart from the singular point $m^*=\mu^*$, the $\zeta$ is parametrically negligible compared to $\eta$ in relativistic cold, degenerate Fermi liquids, as in the non-relativistic case~\cite{SYKES19701}.

Typically, $\tau_\mathrm{rel}$ is proportional to $ T^{-2}$~\cite{Landau2012:physical,PhysRevLett.21.279,SYKES19701,Li:2024iqt}. Consequently, as $T\to 0^+$ the two viscosities behave in opposite ways: $\eta$ diverges while $\zeta$ vanishes. Physically, $\eta$ and $\zeta$ measure resistance to shape and volume deformations, respectively. In the regime near zero temperature, transport is dominated by states in an infinitesimal shell around the Fermi surface~\cite{Delacretaz:2022ocm,Li:2024iqt}. Owing to the conservations of energy and conserved charge number, an isotropic compression or expansion simply rescales the Fermi sphere, producing no energy dissipation. Since the state still remains on an equilibrium manifold, the process is reversible and generates no entropy, hence the bulk channel is inactive and $\zeta=0$. On the other hand, for a scale-invariant fluid, $\zeta$ is strictly zero, as enforced by the form of the stress-energy tensor and the associated Ward identity. A finite $\zeta$, therefore, signals the breaking of scale invariance in the underlying microscopic theory. In some cases, such a breaking is dynamically generated by fluctuations beyond a mean-field description in the non-relativistic system \cite{Enss:2019ydh}. These fluctuations are Pauli-blocked by conservation laws at $T=0$. Shear flow, by contrast, distorts the Fermi sphere into a quadrupolar anisotropy. Relaxing this anisotropy requires momentum-randomizing collisions. However, at $T=0$, elastic isotropic scattering is absent, so the relaxation time diverges. In hydrodynamics, the entropy production density produced in shear channel is $T\partial_\mu s^{\mu}\!\sim\!\eta\,\sigma^{\mu\nu}\sigma_{\mu\nu}$. With the mean free path scaling as $\mathcal{l}\!\sim p_\mathrm{F}^* \tau_\mathrm{rel}/m^*$, set by the effective Fermi momentum $p_\mathrm{F}^*$, we obtain $\mathcal{l}\!\propto T^{-2}$. Hydrodynamic consistency then requires the allowed wave numbers to satisfy $\mathcal{k}\ll \mathcal{l}^{-1}\!\propto T^{2}$. Hence $\sigma^{\mu\nu}\sigma_{\mu\nu}\sim(\mathcal{k}u^\mu)^2$ scales as $T^4$, and the entropy production vanishes as $T\to0^+$. This establishes that bulk dissipation disappears while shear dissipation becomes parametrically suppressed at the edge of hydrodynamic applicability.
\section{Dense nucleon matter at low temperature}\label{sec2}
In general, one of the key applications of Fermi-liquid theory is the description of static nucleon matter. In this work, we adopt the well-known Walecka model~\cite{Walecka:1974qa}, in which the system consists of neutrons (n) and protons (p) interacting through the exchange of scalar and vector mesons. The Lagrangian density is
\begin{align}
    \mathcal{L}_\mathrm{W}=\mathcal{L}_\mathrm{N}+\mathcal{L}_\mathrm{m}+U_\mathrm{m}\,.
\end{align}
The nucleonic part reads
\begin{align}
\label{eq:Lagrangian density1}
    \mathcal{L}_{\mathrm{N}}=\sum_{\rm N=n,p} \bar{\psi}_{\rm N}\left(i \slashed{\partial}-m_{\rm N}+g_{\sigma } \sigma-g_{\omega } \slashed{\omega}+\mu_B\gamma_0\right) \psi_{\rm N}\,,
\end{align}
where $\psi_\mathrm{N}$ denotes the nucleon fields. For simplicity, we assume the isospin symmetry, so neutrons and protons share the same mass $m_\mathrm{N}=939$ MeV and baryon chemical potential $\mu_B$.
\footnote{In what follows, the effective quasiparticle mass and chemical potential will be denoted explicitly for nucleons and baryon.}
The scalar meson field $\sigma$, with mass $m_\sigma=550$ MeV and coupling $g_\sigma=8.69$, generates the long-range attractive interaction, while the vector meson field $\omega^\mu$, with mass $m_\omega=783$ MeV and coupling $g_\omega=8.65$, provides the short-range repulsion. The mesonic dynamics are governed by
\begin{align}
    \label{eq:Lagrangian density2}
    \mathcal{L}_{\rm {m}}=\frac{1}{2}\left(\partial_{\mu} \sigma \partial^{\mu} \sigma-m_{\sigma}^{2} \sigma^{2}\right)-\frac{1}{4} F^{\mu \nu} F_{\mu \nu}+\frac{1}{2} m_{\omega}^{2} \omega^{\mu} \omega_{\mu}\,.
\end{align}
In addition, an effective self-interaction potential
\begin{align}
    \label{eq:Lagrangian density3}
    U_\mathrm{m}=\frac{1}{3} \kappa_3 m_{\rm N}\left(g_{\sigma} \sigma\right)^{3}+\frac{1}{4} \kappa_4\left(g_{\sigma} \sigma\right)^{4}\,,
\end{align}
is introduced to reproduce the empirical saturation properties of nuclear matter with $\kappa_3=7.95\times10^{-3}$ and $\kappa_4=6.95\times10^{-4}$~\cite{Kapusta_Gale_2006}.  

We decompose the meson fields into condensate and fluctuations, $\sigma=\bar{\sigma}+\delta\sigma$ and $\omega^\mu=\bar{\omega}_0+\delta\omega^\mu$, and neglect the fluctuations $\delta\sigma$ and $\delta\omega^\mu$ in the mean-field approximation (MFA). Using the standard imaginary-time path-integral formalism~\cite{Kapusta_Gale_2006}, the effective thermodynamic potential at finite temperature is obtained as~\cite{Li:2022url,Kapusta_Gale_2006}
\begin{align}
    \bar{\mathcal{V}}_\mathrm{W}\left(T,\mu_B\right)=-2p_0\left(T,\mu^*_B, m^*_\mathrm{N}\right)+\frac{1}{2} m_{\sigma}^{2} \bar{\sigma}^{2}-\frac{1}{2} m_{\omega}^{2} \bar{\omega}_{0}^{2}+\bar{U}_\mathrm{m}\,,
\end{align}
where the quasiparticle nucleon mass $m_\mathrm{N}^*$ and effective baryon chemical potential $\mu_B^*$ are shifted by the condensates according to
\begin{align}
    \begin{split}
        m_\mathrm{N}^*\left[\bar\sigma\right]&=m_\mathrm{N}-g_\sigma\bar{\sigma}\left(T,\mu_B\right), \\
        \mu_B^*\left[\bar\omega_0\right]&=\mu_B-g_\omega\bar{\omega}_0\left(T,\mu_B\right),
    \end{split}
\end{align}
and $\bar{U}_\mathrm{m}$ denotes the potential term in MFA.

\begin{figure}[hpbt!]
  \centering
  \includegraphics[width=0.4\textwidth]{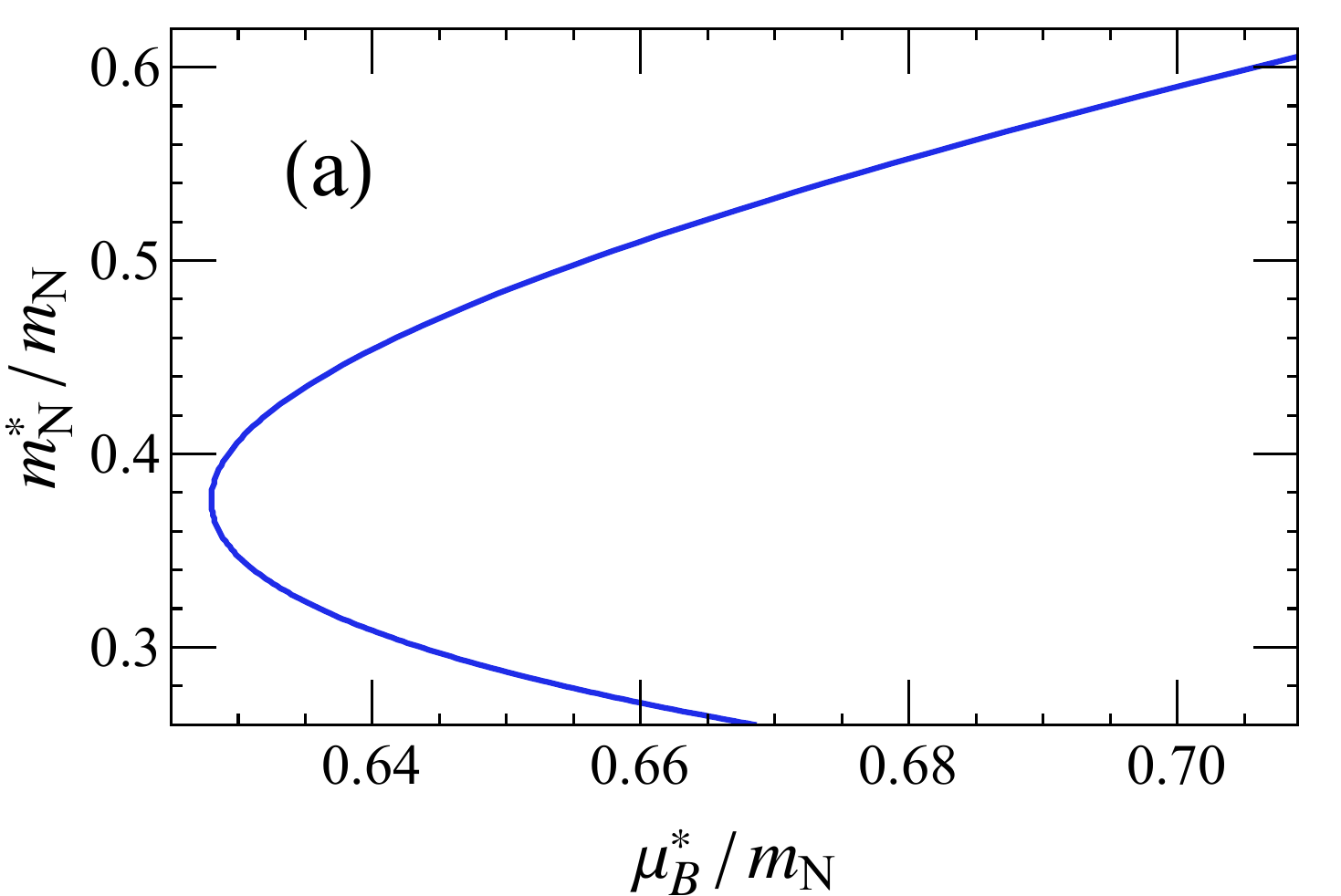}
  \includegraphics[width=0.4\textwidth]{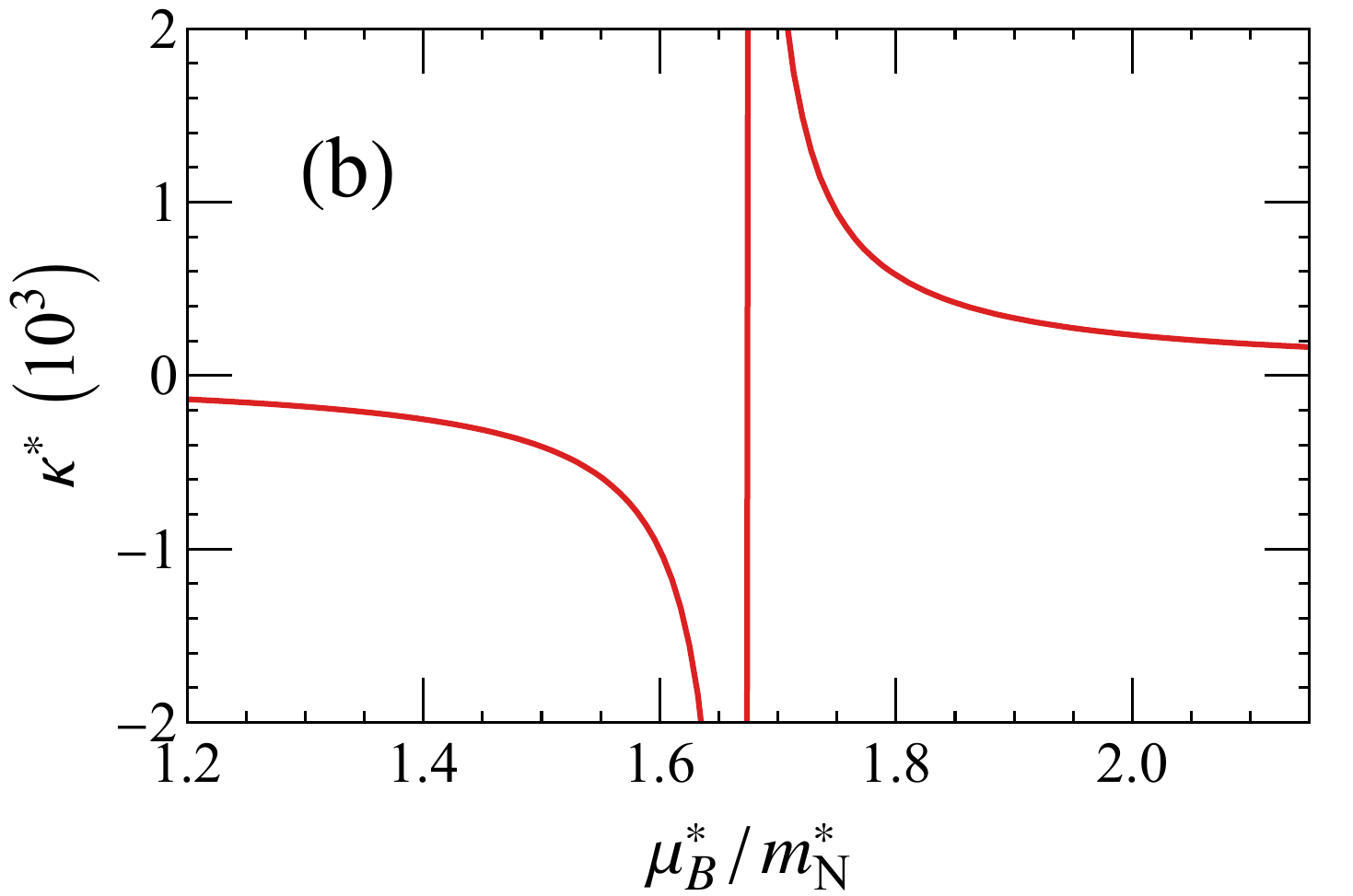}
  \caption{
  Input from the mean-field solution at $T=8$ MeV. (a) Quasinucleon mass $m_\mathrm{N}^*$ and effective baryon chemical potential $\mu_B^*$ from the gap equations Eq.~\eqref{eq:gap_equations}. (b) The response coefficient $\kappa^*$ obtained from Eq.~\eqref{eq:Waleck_dmdmuB}.}
  \label{fig:kappa}
\end{figure}

The pressure of a free fermion species with spin $s$ is
\begin{align}
    \label{eq:thermodynamic-function1}
    p_0(T,\mu,m)=\frac{2s+1}{3\pi^2}\int_{\boldsymbol{p}}\frac{\boldsymbol{p}^2}{E_{\boldsymbol{p}}}\feq\!\left(\boldsymbol{p},\left\{T,\mu,m\right\}\right)\,.
\end{align}
Minimizing $\bar{\cal V}_\mathrm{W}$ leads to the MFA gap equations
\begin{align}
    \label{eq:gap_equations}
	\begin{gathered}
		m_{\omega}^{2} \bar{\omega}_{0}-2 g_{\omega} n_0\left(T,\mu^*_B, m^*_\mathrm{N}\right)=0,\\
		m_{\sigma}^{2} \bar{\sigma}+\frac{\partial \bar{U}_\mathrm{m}}{\partial\bar\sigma}-2 g_{\sigma} n_{\rm s}\left(T,\mu^*_B, m^*_\mathrm{N}\right)=0,
	\end{gathered}
\end{align}
where $n_0$ and $n_{\rm s}$ denote the free particle and scalar densities:
\begin{align}
	\begin{split}
        n_0\left(T,\mu,m\right)&=\left(2s+1\right)\int_{\boldsymbol{p}}\feq\!\left(\boldsymbol{p},\left\{T,\mu,m\right\}\right),\\
        n_{\rm s}\left(T,\mu,m\right)&=\left(2s+1\right)\int_{\boldsymbol{p}}\frac{m}{E_{\boldsymbol{p}}}\feq\!\left(\boldsymbol{p},\left\{T,\mu,m\right\}\right).
	\end{split}
\end{align}
In MFA, the response coefficient $\kappa^*$ can be derived as~\cite{Li:2024iqt}
\begin{align}\label{eq:Waleck_dmdmuB}
    \kappa^*=-\frac{g_\sigma}{g_\omega}\,
    \frac{\partial^2\bar{\mathcal{V}}_\mathrm{W}}{\partial\bar\omega_0\partial\bar\sigma}\,
    \left(\frac{\partial^2\bar{\mathcal{V}}_\mathrm{W}}{\partial\bar\sigma^2}\right)^{-1}.
\end{align}
The sign of $\kappa^*$ cannot be determined from Eq.~\eqref{eq:Waleck_dmdmuB} alone,  although this ambiguity does not affect the positivity of $\zeta$. As noted in Ref.~\cite{Li:2024iqt}, the solution of Eq.~\eqref{eq:gap_equations} corresponds to a saddle point of $\bar{\mathcal{V}}_\mathrm{W}$, which can yield a nonmonotonic $\kappa^*$ with sign changes and apparent divergences (see Fig.~\ref{fig:kappa}). In fact, at finite density, taking $\bar{\mathcal{V}}_\mathrm{W}$ to be imaginary provides a refinement that resolves this artifact~\cite{Nishimura:2016yue,Haensch:2023sig}. A detailed treatment of this point lies beyond the scope of the present work.

Although the fluctuations $\delta\sigma$ and $\delta\omega^\mu$ do not modify the equation of state at MFA, they obey the same dynamics as Eq.~\eqref{eq:Lagrangian density2} at $\mathcal{O}(\delta\sigma^2)$ and mediate nucleon scattering~\cite{Li:2024iqt}. Thus, dissipation arises only beyond the strict MFA. At $T = 8$ MeV, this temperature is insufficient to induce thermal excitations~\cite{ring2004nuclear}. Obtained as the inverse of the LO $2\leftrightarrow2$ scattering rate, $\tau_\mathrm{rel}$ can be parameterized as~\cite{Li:2024iqt}
\footnote{A beyond tree-level scattering calculation based on the Brueckner-Hartree-Fock approach, in which the effective mean free paths relevant for shear viscosity are calculated, is presented in Ref.~\cite{Shternin:2020igy}.}

\begin{align}
	\label{eq:fit_tau}
    \tau_\mathrm{rel}(x)=388\,
    \frac{x^4-2.37x^3+5.44x^2-5.63x+1.90}{x^5+2.25x^4-3.60x^2}\,\mathrm{fm},
\end{align}
with $x=\mu^*_B/m^*_\mathrm{N}$.  

\begin{figure}[htpb!]
	\centering
	\includegraphics[width=0.4\textwidth]{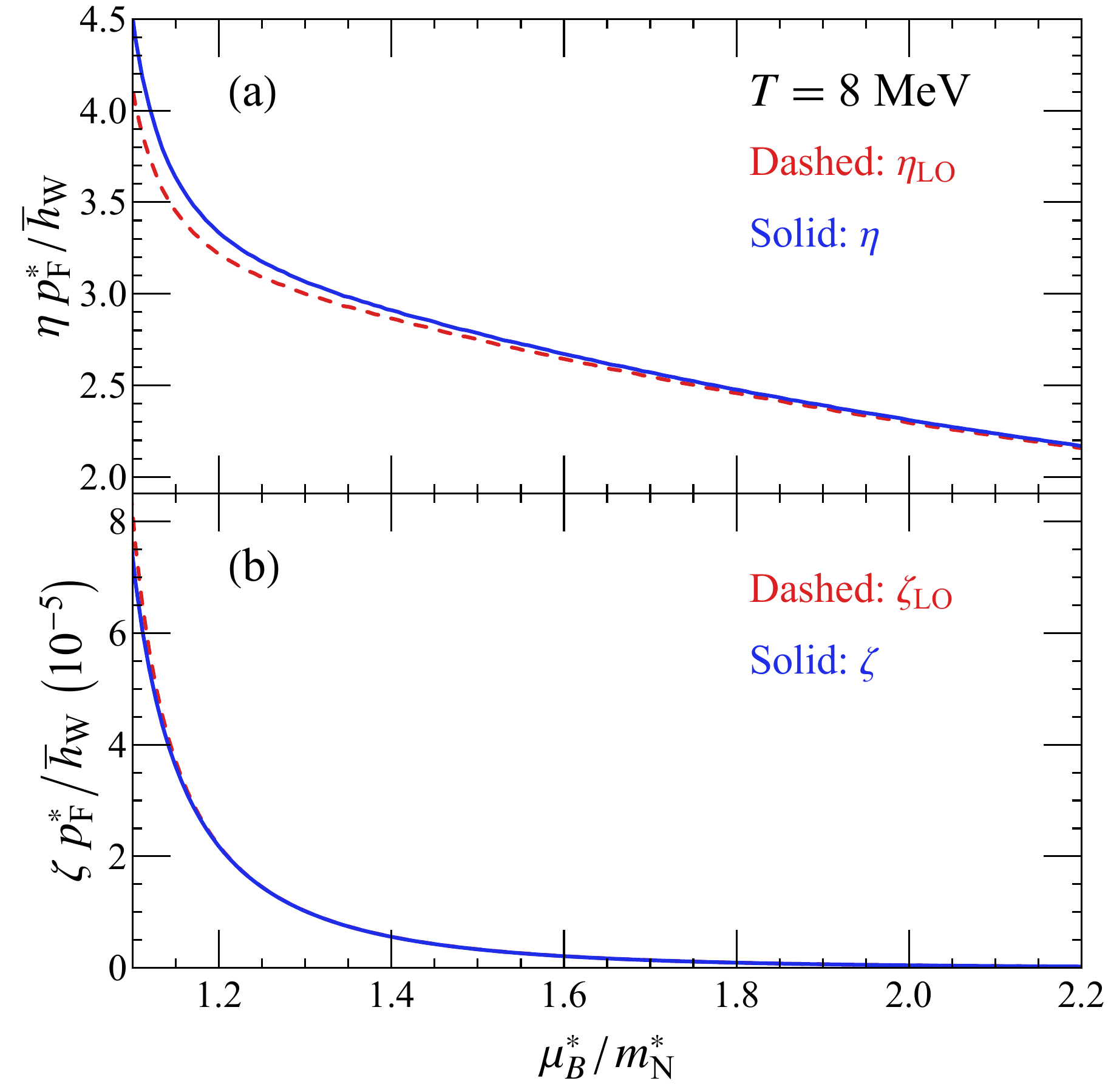}
	\caption{		
		Dimensionless viscosities at $T=8$ MeV. (a) shear viscosity $\eta$ and (b) bulk viscosity $\zeta$, each scaled by the mean-field enthalpy $\bar h_{\mathrm W}$ and effective Fermi momentum $p_\mathrm{F}^*$, shown versus $\mu_B^*/m_\mathrm{N}^*$. Dashed (red) lines are for LO result calculated from Eq.~\eqref{eq:LO_viscosity}. Solid (blue) lines are from full calculations using Eqs.~\eqref{eq:viscosity1}.
        \label{fig:viscosity}
	}
\end{figure}

For dense matter the appropriate measure of fluidity is the enthalpy $h$ rather than the entropy density~\cite{Liao:2009gb,Du:2024wjm}. Within the Walecka model at MFA the enthalpy is
\begin{align}
    \bar{h}_\mathrm{W}\left(T,\mu^*_B, m^*_\mathrm{N}\right)
    =\bar{p}_\mathrm{W}\left(T,\mu^*_B, m^*_\mathrm{N}\right)
    +\bar{\varepsilon}_\mathrm{W}\left(T,\mu^*_B, m^*_\mathrm{N}\right),
\end{align}
where the pressure and energy density are
\begin{align}
    \begin{split}
        \bar{p}_\mathrm{W}&=-\bar{\mathcal{V}}_\mathrm{W},\\
        \bar{\varepsilon}_\mathrm{W}&=2\varepsilon_0\left(T,\mu^*_B, m^*_\mathrm{N}\right)
        +\frac{1}{2} m_{\sigma}^{2}\bar{\sigma}^{2}
        +\frac{1}{2} m_{\omega}^{2}\bar{\omega}_{0}^{2}
        +\bar{U}_\mathrm{m},
    \end{split}
\end{align}
with the free-fermion energy density
\begin{align}
    \varepsilon_0(T,\mu,m)=\left(2s+1\right)\int_{\boldsymbol{p}}E_{\boldsymbol{p}}\,\feq\!\left(\boldsymbol{p},\left\{T,\mu,m\right\}\right).
\end{align}
The resulting dimensionless viscosities, rescaled by $p_\mathrm{F}^*$ and $\bar{\varepsilon}_\mathrm{W}$, are shown in Fig.~\ref{fig:viscosity}.
As anticipated, dissipation in the bulk channel is negligible compared to the shear channel. 
\section{Summary and Outlook}\label{sec3}
In this work, we developed a kinetic framework for cold, dense quasifermion matter. Solving the linearized relativistic Boltzmann equation within RTA tailored to quasiparticles and enforcing Landau matching, we fixed zero-mode ambiguities and proved that the bulk viscosity $\zeta$ is non-negative. A low-temperature analysis yielded compact expressions, $\eta \sim p_{\!F}^{*\,5}\tau_{\rm rel}/\mu^*$ and $\zeta/\eta \propto (T/\mu^*)^4$, establishing that bulk dissipation is parametrically subleading to shear in the degenerate regime. Coupling the framework to the Walecka model at mean field, we presented numerical results over densities relevant to cold, dense nucleon matter, where shear dominates the attenuation of collective motion. We also clarified that apparent pathologies in the bulk channel are artifacts of mean-field stationarity (a saddle point of the effective potential) rather than genuine instabilities of the properly matched hydrodynamics.

These MFA expressions can also be applied to phases with quark and gluon degrees of freedom in the cold and degenerate limit, such as color-superconducting matter \cite{Barrois:1977xd}. Looking ahead, it will be important to replace the RTA by microscopic collision integrals that include in-medium scattering and anisotrpic angular structure, to quantify mesonic fluctuations beyond mean field, and to incorporate isospin asymmetry~\cite{Kaplan:1996xu,Larionov:2021ycq,Fotakis:2022usk,Hu:2022vph}. To sharpen transport inputs for electron-nucleus programs, one should also employ more realistic equations of state and account for the natural deformation of finite nuclei~\cite{Wen:2023oju,Jia:2022ozr}. Finally, cold atom systems offer a complementary arena in which quasiparticle transport can be probed under controlled conditions, providing valuable benchmarks for both hydrodynamic and non-hydrodynamic aspects of the nuclear case~\cite{Brandstetter:2023jsy,Ke:2022tqf,Ke:2023bei,Berges:2025izb}.
\section{Acknowledgment}
We gratefully acknowledge insightful discussions with Michael Buballa, Xu-Guang Huang, Fabian Rennecke, and Yi Yin, fruitful comments from Lorenz von Smekal, as well as editorial assistance from ChatGPT-5. J.L. appreciates the warm hospitality of Fabian Rennecke during his visit to the Institute for Theoretical Physics at Justus-Liebig-Universit\"at Gie\ss{}en. J.L. is supported by the DFG Collaborative Research Centre ``SFB 1225 (ISOQUANT)". J.H. is financially supported by the National Natural Science Foundation of China under Grants No.12505149.
\section{DATA AVAILABILITY}
The data that support the findings of this article are not publicly available. The data are available from the authors upon reasonable request.
\bibliography{Ref}
\end{document}